\documentclass[prd,nofootinbib,showpacs,preprint]{revtex4}
\usepackage{graphicx}
\usepackage{epsfig}
\usepackage{amsmath}
\usepackage{amsfonts}
\usepackage{amssymb}
\usepackage{url}
\usepackage{hyperref}
\usepackage{subfigure}
\usepackage{cancel}
\newcommand{\bd}{\begin{document}}
\newcommand{\ed}{\end{document}}
\newcommand{\bc}{\begin{center}}
\newcommand{\ec}{\end{center}}
\newcommand{\beqa}{\begin{eqnarray}}
\newcommand{\eeqa}{\end{eqnarray}}
\newcommand{\beq}{\begin{equation}}
\newcommand{\eeq}{\end{equation}}
\newcommand{\lsim}{\lesssim}
\newcommand{\gsim}{\gtrsim}
\newcommand{\nn}{\nonumber}
\newcommand{\bmt}{\begin{pmatrix}}
\newcommand{\emt}{\end{pmatrix}}
\usepackage[toc,page]{appendix}
\usepackage{comment}
\def\TeV{\mbox{TeV}}
\def\GeV{\mbox{GeV}}
\def\MeV{\mbox{MeV}}
\def\eV{\mbox{eV}}
\def\keV{\mbox{keV}}
\def\s1{\hat s}
\def\ds{\displaystyle}
\def\lb{\Lambda_b}
\def\ll{\Lambda}
\newcommand{\be}{\begin{equation}}
\newcommand{\ee}{\end{equation}}
\newcommand{\bea}{\begin{eqnarray}}
\newcommand{\eea}{\end{eqnarray}}
\newcommand{\bref}[1]{(\ref{#1})}
\def\slashi#1{\rlap{\sl/}#1}
\newcommand{\<}[1]{\langle {#1} \rangle}
\begin{document}
\title{Study of the rare decays   $B_{s,d}^* \to \mu^+ \mu^-$ }
\author{Suchismita Sahoo }
\author{Rukmani Mohanta }
\affiliation{\,School of Physics, University of Hyderabad,
              Hyderabad - 500046, India  }
\begin{abstract}
 We  investigate the  effect of scalar leptoquarks  and family non-universal $Z^\prime$ boson on  the rare leptonic decays of $B_{s, d}^*$ mesons, mediated
by the FCNC transitions  $b \to (d, s) l^+ l^-$. They are sensitive to a variety of new physics operators as opposed to the $B_{s,d}$ leptonic modes and hence,  can  provide an ideal testing ground to look for new physics  beyond the standard model. We work out the constraint on new physics parameter space using the measured branching ratios of $B_{d, s} \to \mu^+ \mu^-$ processes and the $B_q - \bar B_q$ mixing data. Using the constrained parameters  we estimate the branching ratios of $B_{d,s} \to \mu^+ \mu^-~(e^+ e^-)$ processes. We find that  the branching ratios are reasonably enhanced from their corresponding standard model values in the $X(3,2,1/6)$ leptoquark model and are expected to be  within  the reach of Run II/III of LHC experiments.
\end{abstract}
\maketitle

\section{Introduction}
The standard model (SM) of particle physics is quite  successful  in explaining  almost all the  observed data,
 except the tiny neutrino mass.  Still there are a few reasons to believe that  it  is a
 low  energy  effective theory  and there must be some kind of new physics beyond it.  Furthermore, it  contains some parameters which
are unknown to an adequate level of accuracy. It also does not provide any satisfactory answer to some of the fundamental problems of nature,
such as   the matter dominance of the universe, hierarchy problem, dark matter and dark energy components etc.
Therefore, we need to go beyond the SM to understand some of these open issues.
In this context, the study of the weak decays of $B$ mesons  provides an excellent testing ground to shed light on the nature of new physics
beyond the SM. Recently, several anomalies have been observed at the level of $(3-4)\sigma$   in the rare semileptonic $B$ decays 
mediated through the FCNC transitions $b \to s ll$. These include the well-known angular observable $P_5'$ in $B \to K^* \mu^+ \mu^-$  decay
distribution \cite{lhcb-1a}, 
the lepton flavor non-universality parameter $R_K $ 
in $B \to Kll$ processes \cite{lhcb-1b}, the branching ratio of $B_s \to \phi \mu \mu$ \cite{lhcb-1c},  etc.  Thus,
it is quite natural to expect that, if these deviations are due to the interplay of some kind of new physics, such  effects could also
have an impact in  other observables associated with $b \to s $ transitions. Therefore, it is utmost important to scrutinize as many such observables 
as  possible  to decipher the presence of new physics.

In this regard, some of the most important decay channels are    $B_{s, d} \to l^+ l^-$, 
which are  highly suppressed in the SM as they proceed through  one-loop  penguin and box  diagrams.
Such processes also further suffer from helicity suppression. However,  these processes 
are  theoretically  very clean as the only  hadronic parameter involved is the decay constant of $B$ meson, which can
be precisely known from lattice calculations. The branching ratios of these decay modes are  recently measured by
CMS \cite{cms} and LHCb experiments \cite{lhcb1}  and the corresponding updated  average   values are given by \cite{lhcb2}
\bea
&&{\rm BR}(B_s \to \mu^+ \mu^-)=\left (2.8^{+ 0.7}_{-0.6} \right ) \times 10^{-9},\nn\\
&& {\rm BR}(B_d \to \mu^+ \mu^-)
=\left (3.9_{-1.4}^{+1.6} \right ) \times 10^{-10} \hspace{0.1cm},
\eea
which are almost in agreement with the SM predictions \cite{bobeth1}    
\bea
&&{\rm BR}(B_s \to \mu^+ \mu^-)|_{\rm SM}=\left (3.65 \pm 0.23 \right ) \times 10^{-9},\nn\\
&&{\rm BR}(B_d \to \mu^+ \mu^-)|_{\rm SM}
=\left (1.06 \pm 0.09  \right ) \times 10^{-10}\;.  \label{brmu}
\eea
But certainly they do not rule out the possible existence of NP as the experimental uncertainties are rather large.
Even though there is no direct CP violation observed in these modes but the time dependent oscillation between the $B_s^0$ and $\bar{B}_s^0$
states ($B_s^0-\bar{B}_s^0 $ mixing) may generate the mixing induced CP violation.  The effect of CP violation in  these processes has been studied
in the  literature \cite{bs, bs2} to reveal the signature of new physics.

In this paper, we would like to investigate the  rare leptonic  decay processes $B_{s, d}^* \to \mu^+ \mu^- (e^+ e^-)$, mediated by the
FCNC transitions $b \to s,d$  in the scalar leptoquark (LQ) and 
the family non-universal $Z^\prime$ models.   The vector mesons $B_{s, d}^*$   have the same quark content as the $B_{s, d}$ pseudoscalar mesons  but
their $|\Delta B|=1$ transitions have different sensitivities to the short-distance structure. Furthermore, the 
leptonic decays of $B_{s, d}^*$ mesons are not helicity suppressed and
could have significant branching ratios. These decay modes are recently investigated in the SM in Ref. \cite{grinstein} and here
we are interested to study the effect of new physics in these processes. This in turn  could possibly provide  an alternate way to
 get the hints of new physics from the  observation of these 
processes.    
Experimentally,  only  the radiative decay  modes of these vector mesons i.e.,
$B_{s, d}^* \to B_{s, d} \gamma$ \cite{pdg} transitions are observed.
In the SM,  the branching ratios of $B_{s, d}^* \to \mu^+ \mu^-$ processes are of the order of $\mathcal{O}(10^{-13}-10^{-12})$, if the decay widths of
$B_{s, d}^* \to B_{s, d} \gamma$ modes are considered as $(100-300)$ eV
\cite{bs star decaywidth, bd star decaywidth}. However, for narrower decay widths of $B_{s, d}^* \to B_{s, d} \gamma$ processes, the branching ratios   would be improved, which may not be too far away from
  $B_{s, d} \to \mu^+ \mu^-$ processes.
In this work, we would like to see how the new physics arising from both scalar LQ and family non-universal $Z^\prime$ model affect the branching ratios of these processes.  LQs  are color triplet bosonic particles which couple to quarks and leptons simultaneously and contain both baryon and lepton
quantum numbers. They allow quark-lepton transitions at tree level and thus, explain the quark-lepton universality.
The LQs can have  spin 0 (scalar) or spin 1 (vector) and can  be characterized by their fermion no $(F=3B+L)$ and charge.
Scalar LQs can exist in TeV scale in the extended SM \cite{georgi} such as Grand Unified Theories (GUTs)  \cite{georgi, georgi2},
Pati-Salam model, quark-lepton composite model \cite{kaplan} and the extended technicolor model \cite{schrempp}.
The baryon and lepton number violating  LQs would have  mass near the unification scale to avoid rapid proton decay. 
However, the baryon and lepton number conserving LQs
could be light enough to be accessible in accelerator searches and  they also do not induce
proton decay.
In this work, we consider a simple minimal renormalizable scalar LQ model,
which is invariant under the $SU(3)_C \times SU(2)_L \times U(1)_Y$ gauge group and study the implications of the scalar LQs on the rare leptonic decays of $B_q^*$ mesons.
The effects of scalar LQs in the $B$-sector have been studied extensively in  the literature \cite{Arnold, mohanta1, mohanta2, davidson,leptoquark}.

The $Z^\prime$ boson is a hypothetical color singlet gauge boson, which  could be naturally derived from the extension of  electroweak symmetry of the SM by adding
additional $U(1)^\prime$ gauge symmetry. The $Z^\prime$ gauge boson is predicted in many extended SM theories such as  superstring theories,
grand unified theories, theories with large extra dimensions  and $E_6$ models \cite{E6}.
Among all the relevant $Z^\prime$ models, the family non-universal $Z^\prime$ model \cite{Langacker} is the simplest one to explore the discrepancies
between the observed data  and the corresponding  SM predicted values in some
of the observables associated with $b \to s l^+ l^- $  processes.  It should be noted that the FCNC $b \to s, d$ transitions could be 
induced by family non-universal $U(1)^\prime$ gauge boson
at tree level and can instigate new weak phase, which could  explain the observed CP anomalies in the current experiments. 
Intriguingly the well known $``\pi K$ puzzle"
in the hadronic $B \to \pi K$ decays \cite{Barger3}  and other
anomalies associated with $b \to s \mu \mu$ transitions observed at LHCb could be explained in the $Z^\prime$ model.
The theoretical framework of the  heavy new  $Z^\prime$ gauge boson and its implications in various
rare decay processes  has been investigated in the
literature \cite{Langacker, chang1, Barger, Barger2, z prime, langacker2, Barger3}.
In this paper, we will scrutinize the $Z^\prime$ contribution to the rare $B_{s,d}^* \to \mu^+ \mu^-$ processes with  the allowed parameter space
constrained by $B_q-\bar B_q$ mixing.

The paper is organized as follows. In section II, we  discuss the effective  Hamiltonian describing the $b \to (s, d) l^+ l^-$ transitions and
calculate  branching ratios of  $B_{s,d}^* \to \mu^+ \mu^-$ processes in the standard model. The new physics contribution  to these processes
due to  the scalar LQ exchange  and the constraint on LQ parameter space from $B_{s,d} \to \mu^+ \mu^-$
 are discussed in section III.  In section IV, we obtain 
the branching ratios of $B_{s,d}^* \to \mu^+ \mu^-$ processes  in the family non-universal $Z^\prime$ model using the bounds on $Z^\prime$ couplings from the $B_q-\bar B_q$ mixing. Section V contains our Summary
and Conclusion.

\section{$B_{q}^{(*)} \to \mu^+ \mu^-$ decay processes}
The rare leptonic decays $B_{q} \to l^+ l^-$, mediated by the FCNC transitions $b \to q l^+ l^-~~(q=d, s)$ are  theoretically cleanest $B$ decays.
These decays are strongly suppressed in the SM due to  Glashow-Iliopoulos-Maiani (GIM) mechanism and occur only at one-loop level.
Since the initial particle is a pseudoscalar meson and the final state involves a pair of leptons, these decays encounter additional helicity suppression.
However, the corresponding leptonic decays of vector mesons $B_q^*$ don't suffer from such helicity suppression.
The $B_q^{(*)}$ meson decay constant is the only non-perturbative quantity involved in the description of these processes
which can be reliably calculated using non-perturbative methods such
as QCD sum rules, lattice gauge theory and so on.
The most general effective Hamiltonian
describing $b \to q l^+ l^-$  processes in the SM is  given as \cite{Beneke, kohda}
\bea
{\cal H}_{\rm eff} = - \frac{G_F}{\sqrt 2} \left [\lambda_t^{(q)}{\cal H}_{\rm eff}^{(t)}
+ \lambda_u^{(q)}{\cal H}_{\rm eff}^{(u) }\right ]+ h.c., \label{ham-sm}
\eea
where
\bea
{\cal H}_{\rm eff}^{(u)}&=& C_1(\mathcal{O}_1^c-\mathcal{O}_1^u)+C_2(\mathcal{O}_2^c -\mathcal{O}_2^u), \nn\\
{\cal H}_{\rm eff}^{(t)}&=& C_1 \mathcal{O}_1^c + C_2 \mathcal{O}_2^c + \sum_{i=3}^{10} C_i \mathcal{O}_i
\;.
\eea
 Here  $G_F$ is the Fermi constant  and $\lambda_{q^\prime}^{(q)}=V_{q^\prime b}V_{q^\prime q}^*$
($q^\prime = t,u$) are the product of Cabibbo-Kobayashi-Maskawa (CKM) matrix elements.  The  $C_i$'s are the  Wilson coefficients of the
respective six dimensional operators $\mathcal{O}_i$'s  evaluated at the renormalization scale $\mu = m_b$ \cite{kohda} in the
next-to-next-leading order. The tree level current-current operators $\mathcal{O}_{1, 2}$, the QCD penguin operators $\mathcal{O}_{3,..,6}$  and the
chromo-magnetic operator $\mathcal{O}_8$ do not contribute to the leptonic  processes involving $b \to q ll$ transitions.
They receive contributions only from the electric dipole  operators $\mathcal{O}_{7}^{(\prime)}$ and the semileptonic operators
$\mathcal{O}_{9, 10}^{(\prime)}$, which are defined as
\bea
\mathcal{O}_{7}^{(\prime)} &=& \frac{e}{16 \pi^2} \left[\bar{s}\sigma_{\mu \nu} (m_s P_{L (R)} +m_b P_{R (L)})b\right] F^{\mu \nu}\;,  \nn \\
\mathcal{O}_{9}^{(\prime)} &=& \frac{\alpha}{4\pi} \left(\bar{s} \gamma^\mu P_{L (R)} b\right) \left(\bar{l}\gamma_\mu l\right), \hspace{1cm}
\mathcal{O}_{10}^{(\prime)} = \frac{\alpha}{4\pi} \left(\bar{s} \gamma^\mu P_{L (R)} b\right) \left(\bar{l}\gamma_\mu \gamma_5 l\right),
\eea
where  $P_{L(R)} =(1\mp \gamma_5)/2$ are the projection operators and $\alpha$ denotes the fine structure constant.  The right-handed chiral operators are absent in the SM and can only  be
generated in  various new physics scenarios.
The matrix elements of the quark level operators are related to $B_q^{(*)}$
meson decay constants  as follows:
\bea
\langle 0 | \bar{q}\gamma^\mu  \gamma_5 b | B_q (p_{B_q}) \rangle = -if_{B_q} p_{B_q}^\mu, \nn  \hspace{2.3cm}\\
\langle 0 | \bar{q}\gamma^\mu b | B_q^* (p_{B^*_q}, \varepsilon) \rangle = f_{B_q^*} m_{B_q^*} \varepsilon^\mu, \nn \hspace{2.2cm} \\
\langle 0 | \bar{q}\sigma^{\mu \nu} b | B_q^* (p_{B_q^*}, \varepsilon) \rangle = -i f_{B_q^*}^T  (p_{B_q^*}^\mu \varepsilon^\nu - \varepsilon^\mu p_{B_q^*}^\nu).
\label{matrix}
\eea
Here $\varepsilon^\mu$ is the polarization vector of the $B_q^*$   and $f_{B_q^{(*)}}$  are the decay constant of $B_q^{(*)}$ mesons
which in the heavy quark limit are related as  \cite{decayconstant of bstar}, 
\bea
f_{B_q^*} = f_{B_q}\Big( 1-\frac{2\alpha_s}{3\pi} \Big), ~~~~~~f_{B_q^*}^T = f_{B_q}\Big[ 1+\frac{2\alpha_s}{3\pi}\Big( \log \Big(\frac{m_b}{\mu} \Big)-
1\Big)\Big].
\eea
Now considering the renormalization scale is of the order of mass of $b$ quark ($\mu=m_b$) and neglecting the higher order QCD corrections, 
one can obtain
\bea
f_{B_q^*}=f_{B_q^*}^T  \simeq f_{B_q}.
\eea
 Thus,
the branching ratios of   $B_q \to  \mu^+ \mu^-$ processes in the SM are given as
\bea
{\rm BR}(B_q \to \mu^+ \mu^-) = \frac{G_F^2}{16 \pi^3} \tau_{B_q} \alpha^2 f_{B_q}^2 m_{B_q} m_{\mu}^2 
|V_{tb} V_{tq}^*|^2 \left |C_{10}\right |^2 \sqrt{1- \frac{4 m_\mu^2}{m_{B_q}^2}}.
\eea
Analogously, with Eqns. ({\ref{ham-sm} - \ref{matrix}), the transition amplitudes for $B_q^* \to \mu^+ \mu^-$ processes  in the SM are given by \cite{guang, grinstein} 
\begin{equation}
\mathcal{M} = -\frac{G_F \alpha}{\sqrt{2} \pi} V_{tb} V_{tq}^*  f_{B_q^*} m_{B_q^*} \varepsilon^\mu \Bigg[ \left(C_9^{\rm eff} + 2\frac{m_b}{m_{B_q^*}} C_7^{\rm eff} \right) \left(\bar{\mu} \gamma_\mu \mu\right) + C_{10} \left(\bar{\mu} \gamma_\mu \gamma_5 \mu\right)\Bigg],
\end{equation}
and  the corresponding decay widths as
\begin{equation}
\Gamma (B_q^* \to \mu^+ \mu^-) = \frac{G_F^2 \alpha^2}{96\pi^3} |V_{tb} V_{tq}^*|^2  f_{B_q^*}^2  m_{B_q^*}^2 \sqrt{m_{B_q^*}^2 -4m_l^2} \left[ \Big{|} C_9^{\rm eff} + 2\frac{m_b}{m_{B_q^*}} C_7^{\rm eff} \Big{|}^2 +  \Big{|} C_{10} \Big{|}^2 \right].
\end{equation}
It should be noted that the $B_q^* \to \mu^+ \mu^-$ processes are sensitive to the $C_{7,9}^{\rm eff}$ Wilson coefficients, i.e., ${\cal O}_{7}$ and
${\cal O}_{9}$ operators, whereas the contributions
from these operators vanish  in the case of  corresponding pseudoscalar meson decay processes. The detailed calculation of $B_q^* \to \mu^+ \mu^-$  processes in the SM can be found in \cite{grinstein}.
For numerical calculation, we have taken the particle masses and the CKM matrix elements (in Wolfenstein parametrization)  from \cite{pdg} and  the decay constants of $B_q^{(*)}$ mesons
as $f_{B_s}=225.6\pm 1.1 \pm 5.4$ MeV and $f_{B_s}/ f_{B_d}=1.205 \pm 0.004 \pm 0.007$ from 
Ref. \cite{charles} and obtained the decay rates as
\bea
&&\Gamma(B_s^* \to \mu^+ \mu^-)|_{\rm SM}=\left (1.19 \pm 0.13~ ({\rm CKM}) \pm 0.04 ~({\rm decay~const.}) \right )  \times 10^{-18}~{\rm GeV}, \nn\\
&&\Gamma(B_d^* \to \mu^+ \mu^-)|_{\rm SM}
= \left (3.71 \pm 0.40~ ({\rm CKM}) \pm 0.09 ~({\rm decay~const.}) \right )\times 10^{-20} ~{\rm GeV}\;. 
\eea
In order to compute the branching ratios, we  need to know the total width of $B_{s, d}^*$ bosons.  However, these are
neither measured  nor precisely known   theoretically. Assuming these widths to be coincide with the widths of the flavour-conserving radiative decays, i.e., $\Gamma_{B_{s, d}^*}^{\rm tot} \simeq \Gamma
(B_{s, d}^* \to B_{s, d} \gamma)$, one can write
\bea
\Gamma_{B_{q}^*}^{\rm tot} \simeq \Gamma
(B_{q}^* \to B_{q} \gamma)=\frac{\alpha}{24}|g_{B_{q}^* B_{q} \gamma}|^2
\left (\frac{m_{B_q^*}^2-m_{B_q}^2}{m_{B_q^*}} \right )^3,
\eea
 where the coupling $g_{B_{q}^* B_{q} \gamma}$ is related to the matrix elements of the radiative transitions through
 \bea
 \langle B_q(p) \gamma(q, \epsilon') | B_q^* (p+q, \varepsilon) \rangle=
 \sqrt{4 \pi \alpha}~ g_{B_{q}^* B_{q} \gamma}~ \varepsilon^{\mu \nu \alpha \beta} \epsilon_\mu^{\prime *} q_\nu \varepsilon_\alpha p_\beta\;.
 \eea
These couplings have a simplified parametrization in the heavy hadron chiral perturbation theory (ChPT) \cite{chpt} as
\bea
g_{B_{q}^* B_{q} \gamma} \simeq \frac{Q_b}{m_{B_q^*}}+ \frac{Q_q}{\mu_q}\;,
\label{gqq}
\eea 
where $Q_{b(q)}$ is the charge of the $b$(light quark $q=d,~s$) and $\mu_q$ is a non-perturbative parameter. As discussed in Ref. \cite{mannel},  relation (\ref{gqq}) describes well the measured widths of $D^{*0,+} \to D^{0,+} \gamma$ decays,  provided the parameters $\mu_{u,d}$ would have the value in the range $\mu_{u,d} \simeq 420-430$ MeV. Using the same value of $\mu_{u,d}$, the width for $B_d^*$ meson
is found to be \cite{mannel}
\bea
 \Gamma(B_{d}^* \to B_{d} \gamma) \simeq 0.2~ {\rm KeV}\;.\label{dc1}
\eea
For $B_s^*$ width, using the $SU(3)$-breaking effect to relate $\mu_s$ to 
$\mu_{u,d}$ as  $\mu_s=\mu_{u,d}(m_\rho^2/m_\phi^2)$, one can obtain 
 \bea
 \Gamma(B_{s}^* \to B_{s} \gamma) \simeq 0.07~ {\rm KeV}\;.\label{dc2}
\eea
Recently, using the relativized quark models  Godfrey et al. \cite{godfrey} have obtained these decay widths as 
\bea
 \Gamma(B_{d}^* \to B_{d} \gamma) = 1.23~{\rm KeV}\;, \nn\\
 \Gamma(B_{s}^* \to B_{s} \gamma) = 0.313~ {\rm KeV}\;.
\eea
Since these widths are not yet precisely known, we will present the branching ratios with the values of decay widths as given in Eqns. (\ref{dc1}) and (\ref{dc2}), which can be simply rescaled for any other
values. 
Thus,   the branching ratios of $B_{s,d}^* \to \mu^+ \mu^-$ processes in the SM are
\bea
{\rm BR}(B_s^* \to \mu^+ \mu^-)|_{\rm SM}&=&\left (1.7 \pm 0.21 \right )  
\left (\frac{0.07~{\rm KeV}}{\Gamma_{B_s^*}^{\rm tot}} \right )
\times 10^{-11},\nn \\
{\rm BR}(B_d^* \to \mu^+ \mu^-)|_{\rm SM}
&=&\left (1.86 \pm 0.21 \right )\left (\frac{0.2~{\rm KeV}}{\Gamma_{B_d^*}^{\rm tot}} \right ) \times 10^{-13}\;,
\eea
respectively.   The predicted branching ratios are sizeable and 
 about two orders lower than the branching ratios of  $B_{s, d} \to \mu^+ \mu^-$ processes.

The search for new physics signals in these decay modes would not only be restricted to the branching ratios, but
one can also examine additional observables which are sensitive to new physics. For that purpose 
 we consider the ratio of  various combination of the branching ratios of pseudoscalar and vector $(B_{s, d}^{(*)})$  mesons leptonic decays. 
Recently LHCb has observed $2.6\sigma$ discrepancy in the ratio of $B^+ \to K^+ \mu^+ \mu^-$ to $B^+ \to K^+ e^+ e^-$ branching ratios \cite{Rk}. 
Analogously,  we define the ratio of branching ratios of  $B_q^* \to l^+ l^-$ processes into dimuon over dielectron  as
\begin{equation}
R_{B_q^*} =\frac{BR\left(B_q^* \to \mu^+ \mu^-\right)}{BR\left(B_q^* \to e^+ e^-\right)}= \frac{\sqrt{ 1-4m_\mu^2/m_{B_q^*}^2}}{\sqrt{1 -4m_e^2/m_{B_q^*}^2}},
\end{equation}
which can probe lepton flavour dependent term in and beyond  SM. In the SM, the violation of lepton universality is negligible and $R_{B_q^*}$ is  
around $0.999$ for $B_{s, d}^* $ decays. One can consider another observable, which is the   ratio of decay  rates of $B_{d}^* \to \mu^+ \mu^-$ over 
$B_{s}^* \to \mu^+ \mu^-$ 
processes and its value in the SM  is found to be $0.017$. The measured value of the  analogous observable in the  
pseudoscalar meson case (i.e., 
the ratio of branching ratios of $B_d  \to \mu^+ \mu^-$  to $B_s  \to \mu^+ \mu^-$ decay processes), 
 by the CMS and LHCb  is $0.14^{+0.08}_{-0.06}$ \cite{lhcb2} which has $2.3\sigma$ deviation from the SM prediction 
$0.0295^{+0.0028}_{-0.0025}$ \cite{bobeth1}.
\section{New physics contributions due to scalar leptoquark exchange}

The SM effective Hamiltonian (\ref{ham-sm}) can receive  additional contributions from the scalar LQ exchange.
Here we consider the minimal renormalizable scalar LQ model which are invariant under the SM gauge group
$SU(3)_C \times SU(2)_L \times U(1)_Y$  and conserve baryon number in perturbation theory. There are two such relevant LQ
multiplets $X=(3, 2, 7/6)$ and $(3, 2, 1/6)$, which have sizeable Yukawa couplings to the matter and do  not allow proton decay.
These scalar LQs  potentially contribute to the quark level transitions $b \to (s, d) l^+ l^-$ and
thus, one can constrain the underlying couplings from the experimental measurements on $B_{s, d} \to \mu^+ \mu^-$.
The interaction Lagrangian for $b \to q \mu^+ \mu^-$ $(q=d, s)$ transitions due to  the exchange of  $X(3, 2, 7/6)$ scalar LQs
with the SM fermion bilinear is given by  \cite{Arnold}
\bea
{\cal L}= -\lambda_u^{ij}~ \bar u_{ R}^i X^T \epsilon L^j_L - \lambda_e^{ij}~ \bar e_{ R}^i X^\dagger  Q^j_L + h.c.,\label{lag}
\eea
where  $i, j$ are the generation indices, $X= (V,Y)^T$ is the LQ doublet,  $Q_L$ ($L_L$) denotes the left handed quark (lepton) doublet, 
   the right-handed up-type quark (charged lepton) singlet is represented by $u_R$ ($e_R$)  and $\epsilon = i\sigma_2$ is a $2 \times 2$ matrix. 
Expanding the  $SU(2)$ indices the interaction Lagrangian (\ref{lag}) takes the form 
\bea
{\cal L}= -\lambda_u^{ij}~ \bar u_{\alpha R}^i ( V_\alpha e_L^j - Y_\alpha \nu_L^j )
-\lambda_e^{ij}~ \bar e_R^i \left (V_\alpha^\dagger u_{\alpha L}^j + Y_\alpha^\dagger d_{\alpha L}^j \right )+h.c.\;,\label{lepto}
\eea
which after performing the Fierz transformation and then comparing with the SM effective Hamiltonian (2) yields the  new Wilson coefficients
\bea
C_9^{\rm LQ} = C_{10}^{\rm LQ} = - \frac{ \pi}{2 \sqrt 2 G_F \alpha V_{tb} V_{tq}^* }\frac{\lambda_\mu^{23}{ \lambda_\mu^{2k }}^*}{
M_Y^2}\;,\label{c10np}
\eea
where  $k = 1~ {\rm or}~ 2$ depending on the down type quark $q= d ~{\rm or} ~s$,   $\lambda_{u, e}^{ij}$ are the
LQ couplings analogous to the Yukawa couplings and $Y_\alpha$ and $V_\alpha$ are the LQ fields.

Similarly the interaction Lagrangian for $X(3, 2, 1/6)$ scalar LQ is
 \bea
{\cal L} = - \lambda_d^{ij}~ \bar d^i_{\alpha R} (V_\alpha e_L^j-Y_\alpha \nu_L^j) +h.c.\;,
\eea
which provides the  new Wilson coefficients corresponding to the right-handed chiral (primed) operators $\mathcal{O}_9^\prime$ and $\mathcal{O}_{10}^\prime$ as
\bea
C_9^{'\rm LQ } = - C_{10}^{'\rm LQ } = \frac{ \pi}{2 \sqrt 2 ~G_F \alpha V_{tb}V_{tq}^*} \frac{\lambda_q^{k2} {\lambda_b^{32}}^*}{M_V^2}\;.\label{c10np1}
\eea
Now comparing the SM theoretical predicted values  (\ref{brmu}) with the corresponding  experimental results (1) for the branching ratios of $B_{s, d} \to l^+ l^-$ 
processes, 
one can obtain the  constraints on the new Wilson coefficients $C_{9, 10}^{(\prime)\rm LQ}$ generated in the LQ model. 
In Table I, we present the bounds on the product of LQ couplings   obtained from various $B_{s, d}$ meson decays \cite{mohanta2}. 
Since there exists only the  upper bounds on the branching ratios for $B_{s, d} \to e^+ e^-$ processes, 
the constraints on the corresponding LQ couplings are found to be imprecise.

The decay width of $B_q^* \to \mu^+ \mu^-$ process in the LQ model is
\bea
\Gamma (B_q^* \to \mu^+ \mu^-)&=& \frac{G_F^2 \alpha^2}{96\pi^3} |V_{tb} V_{tq}^*|^2 f_{B_q^*}^2 m_{B_q^*}^2 \sqrt{m_{B_q^*}^2 -4m_l^2} \nn \\
&\times & \left[ \Big{|}(C_9^{\rm eff}+C_9^{\rm LQ}-C_9^{'\rm LQ}) + 2\frac{m_b}{m_{B_q^*}} C_7^{\rm eff} \Big{|}^2   + \Big{|}C_{10}^{\rm SM}+C_{10}^{\rm LQ}-C_{10}^{' \rm LQ}\Big{|}^2 \right]. \; \hspace{0.5cm}
\eea
Now using the values of the new Wilson coefficients from (\ref{c10np}) and (\ref{c10np1}) with the constrained LQ couplings from Table-I, 
the  branching ratios of $B_{s, d}^* \to \mu^+ \mu^- (e^+ e^-)$ processes
both in the $X(3, 2, 7/6)$ and $X(3, 2, 1/6)$ leptoquark model are shown in Table II.  
 From these results, one can see that there is reasonable enhancement
from the SM values of the branching ratios due to the   effect of $X(3, 2, 1/6)$ leptoquark. 
 Even though there is no lepton universality violation  in the SM, the additional LQ particles provide
significant  deviation from the SM and point towards the presence of lepton non-universality in these decays.
The lepton non-universality factor  for $B_s^*$ leptonic decay is found to be  $(0.73-0.999)$ in $X(3, 2, 7/6)$ leptoquark model   
and  $(0.62-0.999)$ for $X(3, 2, 1/6)$ leptoquark model.

We now briefly present the experimental feasibility of these decay modes in the currently running or upcoming
experiments. Because of the large production rate of $b \bar b$ pairs in high-energy $pp$ collisions, the   $B_s^* \to \mu \mu$ mode is
more promising at LHC  compared to Super $B$-factories, where one expects to have not more than $5 \times 10^8$ $B_s^*$ mesons after
$5~{\rm ab}^{-1}$ at $\Upsilon(5S)$ \cite{bell2}. 
As discussed in \cite{grinstein},  let us assume that around 100 $B_s \to \mu \mu$ events will be observed from the Run-I LHC data 
(combined analysis of $3~{\rm fb}^{-1}$ LHCb
and $25~{\rm fb}^{-1}$ CMS). LHC Runs II and III will provide $\sim 10 $ times more data \cite{LHC} and also the production  rate
of $b \bar b$ rate  will be boosted by a factor of 2 due to higher cross section at $\sqrt s=14$ TeV.
Furthermore, after the high-luminosity upgrade of LHC  (HL-LHC), a factor of $\sim 10$ more data is expected. 
Taking into account all these factors, we expect around $\sim 3 \times 10^3 ~(3 \times 10^4)$ $B_s \to \mu \mu$ events by the end of Run III (HL-LHC phase).
Since the branching ratio of $B_s^* \to \mu \mu$ is roughly two order lower than the corresponding $B_s \to \mu \mu$ process, around
30 (300) events are expected to be observed by the end of LHC Run III (HL-LHC). However, if the width of $B_s^* \to B_s \gamma$ is found to be
very narrow, i.e., in the eV range, then the expected number of events will increase.

\begin{table}[htb]
\begin{center}
\caption{Constraints on scalar leptoquark couplings from various leptonic $B_{s,d} \to l^+ l^-$ decays, where $l =e, \mu$.}
\vspace*{0.1 true in}
\begin{tabular}{|c|c|c|}
\hline
Decay Process ~& ~Couplings involved ~&~ Upper bound of  \\
             &  &~the couplings (${\rm GeV^{-2}}$)~  \\
\hline
$B_s \to \mu^\pm \mu^\mp $~~ &~~ $\frac{|\lambda^{32} {\lambda^{22}}^*|}{M_S^2}$ ~~& ~~$ \leq 5 \times 10^{-9} $~\\

\hline

$B_s \to e^\pm e^\mp $ &~ $\frac{|\lambda^{31} {\lambda^{21}}^*|}{M_S^2}$ ~& ~$ < 2.54 \times 10^{-5} $~\\

\hline
$B_d \to \mu^\pm \mu^\mp $ &~ $\frac{|\lambda^{32} {\lambda^{12}}^*|}{M_S^2}$ ~& ~$ (1.5 -3.9 ) \times 10^{-9} $~\\

\hline
$B_d \to e^\pm e^\mp $ &~ $\frac{|\lambda^{31} {\lambda^{11}}^*|}{M_S^2}$ ~& ~$ < 1.73 \times 10^{-5} $~\\

\hline
\end{tabular}
\end{center}
\end{table}


\begin{table}[h]
\caption{The predicted branching ratio  of the  rare $B_{s, d}^* \to l^+ l^-$ decays  in the SM and the LQ model. }
\begin{center}
\begin{tabular}{| c | c | c| c|}
\hline
 Decay process  & Predicted SM Values & Values in $Y=1/6$ LQ model & Values in $Y=7/6$ model \\

 \hline
 \hline
 $B_s^* \rightarrow   \mu^+ \mu^-$  &  $(1.7 \pm 0.2) \times 10^{-11}$& $(1.7-3.19) \times 10^{-11}$& $(1.7-1.93) \times 10^{-11}$ \\

$B_s^* \rightarrow   e^+ e^-$  &  $(1.7 \pm 0.2) \times 10^{-11}$ & $\leq 6.17 \times 10^{-5}$& $\leq 6.17 \times 10^{-5}$ \\

$B_d^* \rightarrow \mu^+ \mu^-$  &  $(1.86 \pm 0.21)\times 10^{-13}$&  $(2.38-8.99)\times 10^{-13}$  &$(2.47-5.4)\times 10^{-13}$\\

$B_d^* \rightarrow   e^+ e^-$  &  $(1.86 \pm 0.21) \times 10^{-13}$& $\leq 6.57 \times 10^{-6}$ & $\leq 6.57 \times 10^{-6}$\\

 \hline
\end{tabular}
\end{center}
\end{table}

\section{$B_{s, d}^* \to \mu^+ \mu^-$ decay process in $Z^\prime$ model}
In the $Z'$ model, the FCNC transitions $b \to q l^+ l^-$, occur at the tree level and the  effective Hamiltonian is given as \cite{Barger2, chang1}
\begin{eqnarray}\label{ZPHbsll}
 {\cal H}_{eff}^{Z^{\prime}}(b\to ql^+l^-)&=&-\frac{2G_F}{\sqrt{2}}
 V_{tb}V^{\ast}_{tq}\left(\frac{g_2 M_Z}{g_1 M_{Z'}}\right)^2\Big[-\frac{B_{qb}^{L}B_{ll}^{L}}{V_{tb}V^{\ast}_{tq}}
 (\bar{q}b)_{V-A}(\bar{l}l)_{V-A} \nn\\
&-&\frac{B_{qb}^{L}B_{ll}^{R}}{V_{tb}V^{\ast}_{tq}}
 (\bar{q}b)_{V-A}(\bar{l}l)_{V+A}\Big]+{\rm h.c.}\,.
\end{eqnarray}
Analogous to the   SM effective Hamiltonian (\ref{ham-sm}), one can  write the Hamiltonian for $Z^\prime$ model  as
\begin{equation}
 {\cal H}_{eff}^{Z^\prime}(b\to ql^+l^-)=-\frac{G_F}{\sqrt{2}}V_{tb}V^{\ast}_{tq}
 \Big[C_{9}^{Z^{\prime}} \mathcal{O}_{9}+C_{10}^{Z^{\prime}} \mathcal{O}_{10}\Big]+{\rm h.c.}\,,
\end{equation}
where the new Wilson coefficients ($C_{9, 10}^{Z^{\prime}}$) are given as
\begin{eqnarray}
 C_{9}^{Z^{\prime}}(M_W)&=&- 2\left(\frac{g_2 M_Z}{g_1 M_{Z'}}\right)^2\frac{B_{qb}^L}{V_{tb}V^{\ast}_{tq}}(B_{ll}^{L}+B_{ll}^{R})\,,\\
 C_{10}^{Z^{\prime}}(M_W)&=& 2\left(\frac{g_2 M_Z}{g_1 M_{Z'}}\right)^2 \frac{B_{qb}^L}{V_{tb}V^{\ast}_{tq}}(B_{ll}^{L}-B_{ll}^{R})\,.
\end{eqnarray}
 Thus,  including these  additional contributions arising from  the  $Z^\prime$ model 
to the  $B_q^* \to \mu^+\mu^-$ processes, the  decay width becomes
\bea
\Gamma (B_q^* \to \mu^+ \mu^-) &=& \frac{G_F^2 }{24\pi^3} |V_{tb} V_{tq}^*|^2 f_{B_q^*}^2 m_{B_q^*}^2 \sqrt{m_{B_q^*}^2 -4m_\mu^2} \nn \\ && 
\Bigg( \Bigg{|}\frac{\alpha}{2\pi}\left(C_9^{\rm eff} + 2\frac{m_b}{m_{B_q}^*} C_7^{\rm eff} \right)- 2\left(\frac{g_2 M_Z}{g_1 M_{Z'}}
\right)^2\frac{B_{qb}^L}{V_{tb}V^{\ast}_{tq}}(B_{\mu \mu}^{L}+B_{\mu \mu}^{R})\Bigg{|}^2  \nn \\ 
&+& \Bigg{|}\frac{\alpha}{2\pi} C_{10}^{\rm SM}+2\left(\frac{g_2 M_Z}{g_1 M_{Z'}}\right)^2 \frac{B_{qb}^L}{V_{tb}V^{\ast}_{tq}}
(B_{\mu \mu}^{L}-B_{\mu \mu}^{R})\Bigg{|}^2 \Bigg) .
\eea
If both the $U(1)$ groups have the same origin from  some grand unified theory, then one can consider $g_2/g_1 \sim 1$. The mass ratio of the SM $Z$ 
and the heavy $Z^\prime$ gauge boson is $M_Z/M_{Z^\prime} \sim 0.1$ for a TeV-scale $Z^\prime$. The chiral  couplings of $Z'$ to leptons 
$(B_{ll}^{L, R})$ are assumed to have the form as the coupling of SM $Z$ boson to leptons  \cite{langacker2}
\bea
B_{ll}^L =T_{3l}^L-\sin^2\theta_W Q_l, ~~~~ B_{ll}^R =T_{3l}^R-\sin^2\theta_W Q_l,
\eea
where $T_{3l}^L~(T_{3l}^R)$ is the third component of weak isospin for the left (right) chiral component  of fermions, 
$Q_l$ is the charge of the fermion and $\theta_W$ is the  weak mixing angle. For all the charged lepton families, $T_{3l}^L=-\frac{1}{2}$ and $T_{3l}^R=0$. 

Next, we need to know the  constraint on $B_{qb}^L$, i.e., FCNC coupling of $Z'$ to  $q$ and $b$ quarks, which can be obtained  
from $B_q-\bar B_q$ mixing parameters, as discussed in the next subsection. 

\subsection{Constraint on $Z^\prime$ couplings from the $ B_q - \bar B_q$ mixing}
In this subsection we estimate the constraints on $Z^\prime$ couplings from the mass difference between the $B_q$-meson mass eigenstates, 
which characterizes the $B_q-\bar{B_q}$ mixing phenomena. Meson-antimeson mixing is sensitive to heavy degrees of freedom that propagate 
in the mixing amplitudes. In the SM, the $B_q-\bar{B_q}$ mixing  occurs through   one-loop level box diagram with   top quark 
and $W$-boson in the loop. The $|\Delta B=2|$ effective Hamiltonian for $ B_q -\bar B_q$ mixing  in the SM is given by \cite{chang1, Buchalla}
\begin{equation}
 \mathcal{H}_{eff}^{SM}(\Delta
 B=2)=\frac{G_F^2}{16\pi^2}M_W^2(V_{tb}V_{tq}^{\ast})^2C^{LL}(\mu_b) \mathcal{O}^{LL}+{\rm h.c.}\,,
\end{equation}
where the operator  $\mathcal{O}^{LL}$ is defined as \cite{Barger}
\begin{eqnarray}
  \mathcal{O}^{LL}
  = [\bar s \gamma_{\mu} (1 - \gamma_5) b]
    [\bar s \gamma^{\mu} (1 - \gamma_5) b] ~, 
\end{eqnarray}
and $C^{LL}$ is the corresponding loop  function.
The $B_q-\bar{B_q}$ mixing amplitude $(M_{12}^{\rm SM})$, corrected up to next-to-leading order (NLO) in QCD is given as \cite{chang1, Barger}
\begin{eqnarray}
\label{M12SM} M_{12}^{\rm SM}(q)&=& \frac{1}{2m_{B_q}} \langle
B_q^0|\mathcal{H}_{\rm eff}^{\rm SM}(\triangle B=2)|\bar{B}_q^0\rangle \nonumber\\
&=&\frac{G_F^2}{12\pi^2}M_W^2(V_{tb}V_{tq}^{\ast})^2(\hat{B}_{B_q}f_{B_q}^2)
m_{B_q} \eta_{B}S_{0}(x_t) \big{[}\alpha_s(\mu_b)\big{]}^{-\frac{\gamma^{(0)}_Q}
 {2\beta_0}}\big{[}1+\frac{\alpha_s(\mu_b)}{4\pi}J_5\big{]}\,,
\end{eqnarray}
where we have used the vacuum insertion method to evaluate the matrix element  as
\bea
\langle \bar{B_q}|\mathcal{O}^{LL}|B_q \rangle = \frac{8}{3} \hat B_{B_q}f^2_{B_q} m^2_{B_q}.
\eea
Here $\hat B_{B_q}$ is the bag parameter, $x_t=(m_t/M_W)^2$ and the ``Inami-Lim"  loop function $S_0(x_t)$ is 
\begin{equation}
  S_0(x_t)
  = \frac{4x_t - 11x_t^2 + x_t^3}{4(1-x_t)^2} - \frac{3x_t^3\ln x_t}{2(1 - x_t)^3} ~.
\end{equation}
 The parameters quoted in (\ref{M12SM}) have values $\gamma^{(0)}_Q=4$, $\beta_0=23/3$ and $J_5=1.627$ \cite{Buchalla}.
The  mass difference between the heavy and light mass eigenstates, which describes the strength of the $ B_q -\bar B_q$ mixing 
is related to the mixing amplitude through 
$\Delta M_q = 2|M_{12}^{\rm SM}|$.  Now using the particle masses from \cite{pdg},
$\eta_B = 0.551$, the bag parameter $\hat{B}_{B_s} = 1.320 \pm 0.017 \pm 0.03$ and $\hat{B}_{B_s}/\hat{B}_{B_d} = 1.023 \pm 0.013 \pm 0.014$  from \cite{charles},
 the mass difference of $\Delta M_s$ and $\Delta M_d$ in the SM are found to be
\bea
&&\Delta M_s^{\rm SM} = \left(17.426 \pm 1.057 \right)~ {\rm ps}^{-1}, \nn \\
&&\Delta M_d^{\rm SM} = \left(0.57 \pm 0.0056 \right) ~{\rm ps}^{-1},
\eea
respectively and their corresponding experimental values are  \cite{pdg}
\bea
\Delta M_s = \left(17.761 \pm 0.022 \right)~{\rm ps}^{-1},  ~~
\Delta M_d = \left(0.51 \pm 0.003 \right) ~{\rm ps}^{-1}.
\eea
Although there is no noticeable difference  in the mass difference between the theoretical predictions and the corresponding experimental values,
the ratio of the experimental and SM values are found  to be
\bea
\Delta M_s/\Delta M_s^{\rm SM} = \left(1.019 \pm 0.062 \right),  ~~
\Delta M_d/\Delta M_d^{\rm SM} = \left(0.895 \pm 0.01 \right).
\eea
We use these values to constrain the new physics parameter space of the family non-universal $Z'$ model.
In this model the $B_q -\bar B_q$ mixing can occur at tree level and the corresponding effective Hamiltonian
is given as \cite{Barger}
\begin{equation}
  \label{eq:para}
  {\cal H}_{\rm eff}^{Z^\prime} = \frac{G_F}{\sqrt{2}} \left( \frac{g_2
      M_Z}{g_1 M_{Z'}} B_{qb}^L \right)^2  \mathcal{O}^{LL}(m_b) \equiv
  \frac{G_F}{\sqrt{2}} (\rho_q^L)^2 e^{2 i \phi_q^L} \mathcal{O}^{LL}(m_b)~,
\end{equation}
where  $B_{qb}^L$ is the flavour-off-diagonal left handed FCNC $b_L-q_L-Z^\prime$ couplings to the bottom and other down type quark and
$M_{Z^\prime}$ is the mass of the new $Z^\prime$ gauge boson. Here $g_2$ and $g_1$ 
are the gauge couplings of $Z^\prime$ and $Z$ bosons respectively and $g_1=e/(\sin\theta_W \cos\theta_W)$.
The parameter $\rho_q^L$ is defined as
\bea
\rho_q^L = \frac{g_2 M_Z}{g_1 M_{Z'}} B_{qb}^L,
\eea
and $\phi_q^L$ is the weak phase in the $Z^\prime$ model. 
 Here we ignore the $Z-Z^\prime$ mixing for simplicity and assume that there is no remarkable renormalization group evolution effects 
between the $M_{Z^\prime}$ and $M_W$ scale.  
After RG evolution from $M_W$ scale to $m_b$ scale, the contribution due to additional $Z^\prime$ gauge boson exchange 
to the mass difference $M_{12}^{Z^{\prime}}$ is given by  \cite{chang1, Barger}
\begin{equation}
\label{M12ZP}
M_{12}^{Z^{\prime}}(q)=\frac{G_F}{2\sqrt{2}}
 |\rho_{q}^{L}|^2e^{i2\phi_q^L}\frac{8}{3}m_{B_q}(\hat{B}_{B_q}f_{B_q}^2)\big{[}\alpha_s(\mu_W)/\alpha_s(\mu_b)\big{]}^{\frac{\gamma^{(0)}_Q}
 {2\beta_0}}\big{[}1+\frac{\alpha_s(\mu_b)-\alpha_s(\mu_W)}{4\pi}J_5\big{]}\,.
\end{equation}
Thus, including both SM and $Z^\prime$ couplings, the total contributions to the mass difference is given by
\bea
\Delta M_q &=& \Delta M_q^{\rm SM}+\Delta M_q^{Z^{\prime}} \nn \\
           &=& \Delta M_q^{SM} \left|1+\frac{8\sqrt{2}\pi^2 U^\prime_{LL}}{M_W^2 G_F |V_{tb}V_{tq}^*|^2 \eta_B  S_0(x_t)} (\rho_q^L)^2 e^{i 2\phi_q^L} \right|\;,
\eea
with
\begin{equation}
 U_{LL}^{\prime}\equiv\big{[}\alpha_s(\mu_W)\big{]}^{\frac{\gamma^{(0)}_Q}
 {2\beta_0}}\big{[}1-\frac{\alpha_s(\mu_W)}{4\pi}J_5\big{]}.
\end{equation}
The constraints on $\rho_q^L, ~~(q=s, d)$ parameter space can be obtained by varying the ratio of mass difference $(\Delta M_q/\Delta M_q^{SM})$ 
within its $2\sigma$ allowed range as shown in Fig. 1. Here the left plot represents the constraint due to the $B_s-\bar{B_s}$ mixing and the 
right plot for the  $B_d-\bar{B_d}$ mixing. From these plots,  
the constraint on $\rho_s^L$  for the entire range of $\phi_s^L$ in  $B_s-\bar{B_s}$ mixing is found to be
\bea
0 \leq \rho_s^L \leq 0.5 \times 10^{-3} ~~{\rm for}~~0\leq \phi_s^L \leq \pi \;.\label{bs-bar}
\eea
Similarly for $B_d-\bar{B_d}$ mixing case the bound is
\bea
1 \times 10^{-4}  \leq \rho_d^L \leq 1.25 \times 10^{-4} ~~{\rm for}~~\pi/3 \leq \phi_d^L \leq 2\pi/3. \label{bd-bar}
\eea
\begin{figure}[h]
\centering
\includegraphics[scale=0.65]{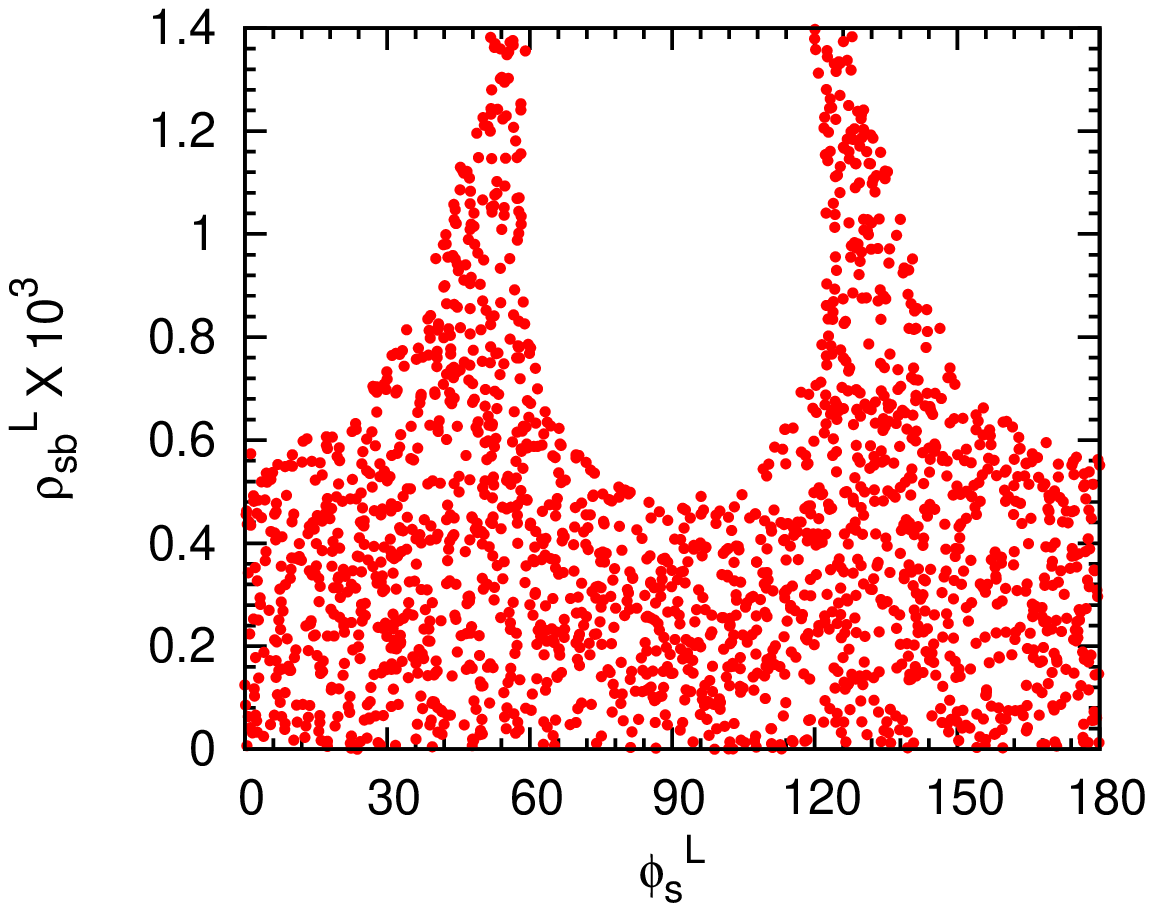}
\quad
\includegraphics[scale=0.65]{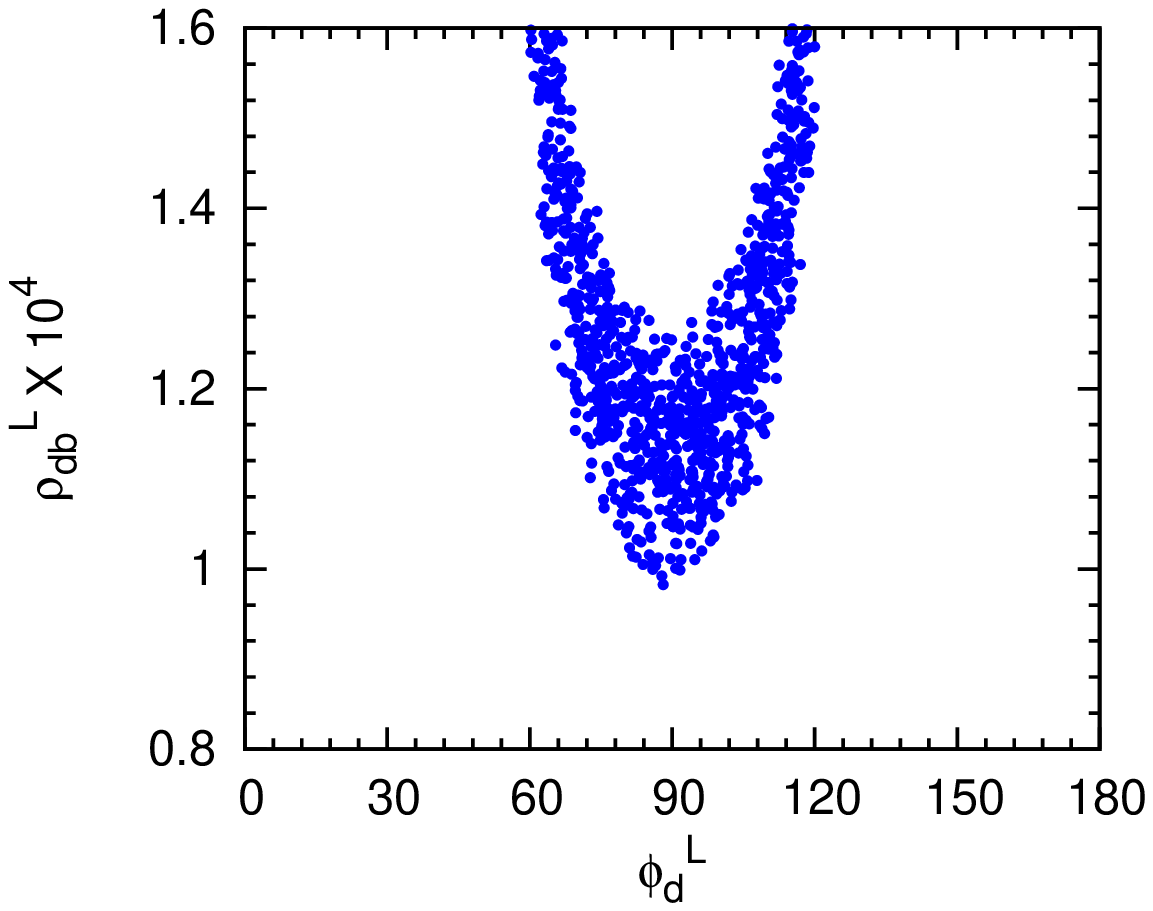}
\caption{The allowed region of $\rho_q^L$ and $\phi_q^L$ obtained from the mass difference  between $B_q$-meson mass eigenstates. The left panel corresponds to the constraints from the $B_s-\bar{B_s}$ mixing  and right panel is for $B_d-\bar{B_d}$ mixing.}
\end{figure}
After knowing the constraints on the $Z^\prime$ couplings to quarks $(B_{qb}^L)$, we now  proceed to calculate the branching ratios for
 $B_q^* \to \mu^+ \mu^-$ processes.
Using the values from  (\ref{bs-bar}) and  (\ref{bd-bar}),  the branching ratios for $B_{s, d}^* \to \mu^+ \mu^-$ processes  are found to be
\bea
{\rm BR}(B_s^* \to \mu^+ \mu^-)|_{Z^\prime}&=&\left (1.7-2.2 \right )\left( \frac{ 0.07~{\rm KeV}}{\Gamma_{B_s^*}}\right ) \times 10^{-11},\nn \\
{\rm BR}(B_d^* \to \mu^+ \mu^-)|_{Z^\prime}
&=&\left (1.67-2.23 \right ) \left (\frac{ 0.2~{\rm KeV}}{\Gamma_{B_d^*}}\right ) \times 10^{-13}.\;
\eea
The effect of $Z^\prime$ boson to the branching ratios of leptonic decays  $B_q^* \to \mu^+ \mu^-$  is 
very marginal, and  the predicted branching ratios are almost 
comparable to the   corresponding  SM values.  

\section{Conclusion}
In this paper, we have studied the  pure leptonic  decay processes of $B_{s, d}^*$ vector mesons in  
the scalar leptoquark  and family non-universal $Z^\prime$ models 
and  estimated the branching ratios of  $B_{s, d}^* \to  \mu^+ \mu^-~( e^+ e^-)$ processes.  These decays are not 
chirally suppressed and sensitive to the semileptonic operator $\mathcal{O}_7,~\mathcal{O}_9$ and $\mathcal{O}_{10}$,  which could provide information 
of new physics at the TeV scale. These decay modes are recently studied in the SM in Ref. \cite{grinstein}.  
The SM  branching ratios of $B_{s, d}^* \to  \mu^+ \mu^-$ processes are found to be of the 
order of $\mathcal{O}(10^{-11})$/$\mathcal{O}(10^{-13})$,
 which are roughly two order lower than the  
$B_{s, d} \to  \mu^+ \mu^-$ processes.
So these modes are expected to be observed in the Run III of LHC experiments.
For the LQ sector, we consider both 
the $X(3,2,7/6)$ and $X(3,2,1/6)$  relevant  LQ models. The  leptoquark parameter space is constrained by using the branching 
ratios of $B_{s, d} \to \mu^+ \mu^-$ $(e^+ e^-)$  processes and the $Z^\prime$ couplings are constrained by 
the $B_q-\bar B_q$ mixing parameters. The $X(3,2,1/6)$ LQ  provides significant enhancement to
 $B_{s, d}^* \to \mu^+ \mu^-$ processes in comparison to the $X(3,2,7/6)$ LQ and $Z^\prime$ models.  
The   observation  of these decay modes in LHC experiments    will definitely shed light on the nature 
 of new physics beyond the SM.


{\bf Acknowledgments}

We would like to thank Science and Engineering Research Board (SERB),
Government of India for financial support through grant No. SB/S2/HEP-017/2013.


\end{document}